\documentclass[preprint,12pt]{elsarticle}

\usepackage{mdframed}
\usepackage{listings}
\usepackage{url}
\usepackage{amsmath,amsfonts,amssymb}
\usepackage[ruled,vlined]{algorithm2e}
\usepackage{lipsum}
\usepackage{amsfonts}
\usepackage{graphicx}
\usepackage{epstopdf}
\usepackage{algorithmic}
\usepackage{caption}
\usepackage{fontawesome}
\ifpdf
  \DeclareGraphicsExtensions{.eps,.pdf,.png,.jpg}
\else
  \DeclareGraphicsExtensions{.eps}
\fi
\usepackage[
    a4paper,%
    bindingoffset=0.0in,%
    left=0.8in,%
    right=0.8in,%
    top=1.in,%
    bottom=1.in,%
    footskip=.4in]{geometry}

\usepackage{url}
\usepackage{color}
\usepackage[dvipsnames]{xcolor}
\usepackage{xspace}
\usepackage{numprint}
\usepackage{amsopn}
\usepackage{float}
\usepackage[overload]{empheq}
\usepackage{upgreek}
\usepackage{physics}
\usepackage{subfig}

\usepackage{multirow}
\usepackage{ctable}
\usepackage{rotating}
\usepackage{lscape}
\usepackage{array}
\usepackage{booktabs}
\usepackage{supertabular}
\usepackage{tabularx}
\usepackage{hyperref}

\newcommand{\ie}{\textit{i}.\textit{e}., }

\newcounter{bla}

\journal{Computer Physics Communications}

\begin{document}

\begin{frontmatter}



\title{Unlocking massively parallel spectral proper orthogonal decompositions in the PySPOD package}


\author[a]{Marcin Rogowski\corref{author}}
\author[b]{Brandon C. Y. Yeung}
\author[b]{Oliver T. Schmidt}
\author[f]{Romit Maulik}
\author[a]{Lisandro Dalcin\corref{author}}
\author[a,c]{Matteo Parsani}
\author[d,e]{Gianmarco Mengaldo\corref{author}}

\cortext[author] {Corresponding authors.\\\textit{E-mail addresses: marcin.rogowski@kaust.edu.sa, \; dalcinl@gmail.com, \; mpegim@nus.edu.sg}}
\address[a]{King Abdullah University of Science and Technology (KAUST), Computer Electrical and Mathematical Science and Engineering Division (CEMSE), Extreme Computing Research Center (ECRC), Thuwal, Saudi Arabia}
\address[b]{Department of Mechanical and Aerospace Engineering, University of California San Diego, La Jolla, CA, USA}
\address[c]{King Abdullah University of Science and Technology (KAUST), Physical Science and Engineering Division (PSE), Thuwal, Saudi Arabia}
\address[d]{Department of Mechanical Engineering, National University of Singapore, Singapore, SG}
\address[e]{Honorary Research Fellow, Department of Aeronautics, Imperial College London, London, UK}
\address[f]{Information Science and Technology Department, Pennsylvania State University, State College, PA, USA}

\begin{abstract}
We propose a parallel (distributed) version of the spectral proper orthogonal decomposition (SPOD) technique. The parallel SPOD algorithm distributes the spatial dimension of the dataset preserving time. This approach is adopted to preserve the non-distributed fast Fourier transform of the data in time, thereby avoiding the associated bottlenecks. The parallel SPOD algorithm is implemented in the \href{https://github.com/mathe-lab/PySPOD}{PySPOD} library and makes use of the standard message passing interface (MPI) library, implemented in Python via \href{https://mpi4py.readthedocs.io/en/stable/}{mpi4py}. An extensive performance evaluation of the parallel package is provided, including strong and weak scalability analyses. The open-source library allows the analysis of large datasets of interest across the scientific community. Here, we present applications in fluid dynamics and geophysics, that are extremely difficult (if not impossible) to achieve without a parallel algorithm. This work opens the path toward modal analyses of big quasi-stationary data, helping to uncover new unexplored spatio-temporal patterns.





\end{abstract}


\end{frontmatter}



{\bf PROGRAM SUMMARY/NEW VERSION PROGRAM SUMMARY}

\begin{small}
\noindent
{\em Program Title:}  PySPOD \\
{\em CPC Library link to program files:} (to be added by Technical Editor) \\
{\em Developer's repository link:} \url{https://github.com/MathEXLab/PySPOD} \\
{\em Code Ocean capsule:} (to be added by Technical Editor)\\
{\em Licensing provisions(please choose one):} MIT License \\
{\em Programming language:} Python \\
{\em Nature of problem:} Large spatio-temporal datasets may contain coherent patterns that can be leveraged to better understand, model, and possibly predict the behaviour of complex dynamical systems. To this end, modal decomposition methods, such as the proper orthogonal decomposition (POD) and its spectral counterpart (SPOD), constitute powerful tools. The SPOD algorithm allows the systematic identification of space-time coherent patterns. This can be used to understand better the physics of the process of interest, and provide a path for mathematical modeling, including reduced order modeling. The SPOD algorithm has been successfully applied to fluid dynamics, geophysics and other domains. However, the existing open-source implementations are serial, and they prevent running on the increasingly large datasets that are becoming available, especially in computational physics. The inability to analyse via SPOD large dataset in turn prevents unlocking  novel mechanisms and dynamical behaviours in complex systems. \\
{\em Solution method: } We provide an open-source parallel (MPI distributed) code, namely PySPOD, that is able to run on large datasets (the ones considered in the present paper reach about 200 Terabytes). The code is built on the previous serial open-source code PySPOD that was published in \url{https://joss.theoj.org/papers/10.21105/joss.02862.pdf}. The new parallel implementation is able to scale on several nodes (we show both weak and strong scalability) and solve some of the bottlenecks that are commonly found at the I/O stage. The current parallel code allows running on datasets that was not easy or possible to analyse with serial SPOD algorithms, hence providing a path towards unlocking novel findings in computational physics. \\
{\em Additional comments including Restrictions and Unusual features (approx. 50-250 words):} The code comes with a set of built-in postprocessing tools, for visualizing the results. It also comes with extensive continuous integration, documentation, and tutorials, as well as a dedicated website in addition to the associated GiHub repository. Within the package we also provide a parallel implementation of the proper orthogonal decomposition (POD), that leverages the I/O parallel capabilities of the SPOD algorithm. \\
\end{small}

\section{Introduction}\label{sec:introduction}
Data that depends on both space and time, also referred to as spatio-temporal data, is ubiquitous. It can represent the Earth's atmosphere, the flow past an aircraft, and the ocean's dynamics, among many other phenomena and processes. Usually, spatio-temporal data is high-dimensional and its interpretation non obvious. Yet, spatio-temporal data, especially the one related to physical processes, contain coherent patterns. These, if uncovered, may provide a better understanding of critical aspects of the underlying physical processes of interest. Hence, tools to analyze and make sense of this type of data are of paramount importance. 

Over the last several years, many tools have been proposed to mining information from spatio-temporal data, such as the proper orthogonal decomposition (POD)~\cite{berkooz1993proper,maulik2021pyparsvd}, the dynamic mode decomposition (DMD)~\cite{schmid2010dynamic}, and the spectral proper orthogonal decomposition (SPOD)~\cite{lumley1967,towne2018spectral,schmidt2019spectral,schmidt2020guide,lario2022neural}. Yet, the majority of these tools comes with limited parallel capabilities for handling large datasets. 

In this paper, we propose a first parallel (distributed) SPOD algorithm, that allows large-scale modal decomposition analyses. The algorithm has been tested on both geophysical and fluid mechanics data, reaching 199~terabytes (TB) of size. Weak and strong scalability have been thoroughly assessed for the entire algorithm as well as for its main constituting steps, including  input/output handling, discrete Fourier transform, and eigenvalue computations. The efficiency of the novel parallel algorithm proposed in this paper allows analysis of data at unprecedented scales, opening the path towards uncovering new physics in large datasets, that existing SPOD packages were unable to tackle \cite{schmidt2022spectral,Mengaldo2021}.

This paper is organized as follows. In section~\ref{sec:spod_method}, we introduce the SPOD method. In section~\ref{sec:parallel-spod}, we detail our parallelization strategy. In section~\ref{sec:datasets}, we present the results obtained on two large datasets, one related to fluid mechanics and the other to geophysics (in particular atmospheric physics). In section~\ref{sec:scalability}, we present the strong and weak scalability analysis of the parallel SPOD algorithm. In section~\ref{sec:conclusions}, we provide some concluding remarks.

\section{The spectral proper orthogonal decomposition}\label{sec:spod_method}

\subsection{A note on suitable data}\label{sec:data_suitability}
The spectral proper orthogonal decomposition, also referred to as spectral POD or simply SPOD, extracts coherent structures from statistically stationary data. The data considered have spatial and temporal dependence, and describe a stochastic (also referred to as random) process denoted by $\boldsymbol{q}(\boldsymbol{x}, t)$, where $\boldsymbol{x}$ represents the spatial coordinate, and $t$ the time. Usually, $\boldsymbol{x} = (x,y,z) \in \mathbb{R}^3$ for the processes of interest, where $x, y$, and $z$ are Cartesian coordinates. In practice, both two- and three-dimensional data may be used. We now define in more detail the concept of stationarity and the assumptions that allow us to use ensemble and time averages interchangeably. 

The \textit{stationarity} assumption under which the SPOD operates is typically intended in the \textit{weak sense}, also referred to as \textit{wide-sense} or \textit{covariance stationarity}. This implies that $\boldsymbol{q}(\boldsymbol{x}, t)$, where we dropped the probability parametrization $\xi$, has first- and second-order moments (\ie average and autocovariance) that do not vary with time. The mathematical formalization of wide-sense stationarity is as follows. Let $\boldsymbol{q}(\boldsymbol{x}, t)$ be a continuous-time stochastic process, $E[\cdot]$ be the expectation operator, and $\boldsymbol{\mathcal{C}}$ be the covariance. If 
\begin{itemize}
\item the expectation operator is independent of $t$:
\begin{equation}
E[\boldsymbol{q}(\boldsymbol{x}, t)] = \mu(\boldsymbol{x}),
\label{eq:mean-indep}
\end{equation}
\item the covariance depends only on the difference between two times, $t - t'$:
\begin{equation}
\boldsymbol{\mathcal{C}}(\boldsymbol{x}, \boldsymbol{x}', t, t') = \boldsymbol{\mathcal{C}}(\boldsymbol{x}, \boldsymbol{x}', \tau), \quad \text{where $\,\tau = t - t'$}, 
\label{eq:cov-indep}
\end{equation}
\item the average `power' is bounded, and does not go to infinity:
\begin{equation}
E[|\boldsymbol{q}(\boldsymbol{x},t)^2|] < \infty,
\label{eq:energy-bounded}
\end{equation}
\end{itemize}
then $\boldsymbol{q}(\boldsymbol{x}, t)$ is said to be \textit{wide-sense stationary}. 

Many fields in computational physics, including fluid dynamics and geophysics, give rise to wide-sense stationary problems. In section~\ref{sec:datasets} we will apply SPOD to examples of wide-sense stationary problems in these two disciplines. 

As a final yet important note in this section, we add that the stochastic processes considered are also ergodic in addition to being wide-sense stationary. This means that the expectation, $E[\boldsymbol{q}]$, coincides with the ensemble average of different realizations of $\boldsymbol{q}(\boldsymbol{x},t)$, that in turn is equal to the long-time average, $\overline{\boldsymbol{q}(\boldsymbol{x}, t)}$. Therefore, when we say zero-average stochastic process, we refer to a process where $\bar{\boldsymbol{q}}=E[\boldsymbol{q}]$ has been removed. Removal of the mean facilitates the interpretation of the SPOD eigenvalues as perturbation energy or variance. In the following, without loss of generality we will always assume that $\bar{\boldsymbol{q}} = 0$.  In sections~\ref{sec:theory} and \ref{sec:discrete}, we will use some of the notions introduced here to derive the continuous and discrete SPOD approaches.

\subsection{Theory}\label{sec:theory}
For the sake of reader's convenience, we report here the theory behind SPOD, closely following~\cite{towne2018spectral}. 

The task of the SPOD is to identify a deterministic function $\boldsymbol{\phi}(\boldsymbol{x},t)$ (or a set of functions) that best approximates the weak-sense stationary and zero-average process $\boldsymbol{q}(\boldsymbol{x},t)$. In mathematical terms, this translates into finding the function $\boldsymbol{\phi}(\boldsymbol{x},t)$ that maximizes the expected value of the normalized projection of the stochastic function $\boldsymbol{q}(\boldsymbol{x},t)$, that is,
\begin{equation}\label{eq:spod_statement}
\lambda = \frac{E\big[|\langle \boldsymbol{q}(\boldsymbol{x},t),\boldsymbol{\phi}(\boldsymbol{x},t) \rangle_{\boldsymbol{x},t}|^2\big]}{\langle \boldsymbol{\phi}(\boldsymbol{x},t),\boldsymbol{\phi}(\boldsymbol{x},t) \rangle_{\boldsymbol{x},t}}.
\end{equation}
In equation~\eqref{eq:spod_statement}, we assume that any realization of $\boldsymbol{q}(\boldsymbol{x},t)$ belongs to a Hilbert space, $H$, with a space-time inner product, $\langle \cdot,\cdot \rangle_{\boldsymbol{x},t}$, and expectation operator, $E[\cdot]$, here taken to be the ensemble average. The inner product in equation~\eqref{eq:spod_statement}, $\langle \cdot,\cdot \rangle_{\boldsymbol{x},t}$, between two generic variables, $\boldsymbol{u}$ and $\boldsymbol{v}$, is defined as
\begin{equation}\label{eq:innerprod_spacetime}
\langle \boldsymbol{u},\boldsymbol{v} \rangle_{\boldsymbol{x},t} = \int_{-\infty}^{\infty} \int_{\Omega} \boldsymbol{u}^*(\boldsymbol{x},t)\; \boldsymbol{W}(\boldsymbol{x})\; \boldsymbol{v}(\boldsymbol{x},t) \, \text{d}\boldsymbol{x}  \, \text{d}t, 
\end{equation}
where $\Omega$ denotes the spatial domain, $\boldsymbol{W}(\boldsymbol{x})$ the spatial weighting, and the asterisk superscript represents the conjugate transpose. By invoking the Karhunen-Lo{\'e}ve (KL) theorem~\cite{kosambi2016statistics,loeve2017probability}, we know that there exists a set of mutually orthogonal deterministic functions that form a complete basis in $H$. This can be defined as $\hat{\boldsymbol{q}}(\boldsymbol{x},f) = \sum_{k=1}^{\infty} a_k(f)\boldsymbol{\phi}_{k}(\boldsymbol{x},f)$, where $\hat{(\cdot)}$ denotes the Fourier transform in time. The eigenfunctions, $\boldsymbol{\phi}_{k}$, and their associated eigenvalues, $\lambda_{k}$, are solutions to the eigenvalue problem in the frequency domain:
\begin{equation}\label{eq:spod}
\int_{\Omega} \hat{\boldsymbol{\mathcal{C}}}(\boldsymbol{x},\boldsymbol{x}',f) \, \boldsymbol{W}(\boldsymbol{x}') \, \boldsymbol{\phi}(\boldsymbol{x}',f) \, \text{d}\boldsymbol{x}'  = \lambda(f) \boldsymbol{\phi}(\boldsymbol{x},f),
\end{equation}
where $\hat{\boldsymbol{\mathcal{C}}}(\boldsymbol{x},\boldsymbol{x}',f)$ is the cross-spectral density tensor, i.e., the Fourier transform of $\boldsymbol{\mathcal{C}}$:
\begin{equation}\label{eq:csd}
\hat{\boldsymbol{\mathcal{C}}}(\boldsymbol{x},\boldsymbol{x}',f) = \int_{-\infty}^{\infty} \boldsymbol{\mathcal{C}}(\boldsymbol{x},\boldsymbol{x}',\tau) e^{-\mathrm{i}2\pi f \tau} \text{d}\tau.
\end{equation}
In equation~\eqref{eq:csd}, the two-point space-time correlation tensor  $\boldsymbol{\mathcal{C}}(\boldsymbol{x},\boldsymbol{x}',\tau)$ is defined as 
\begin{equation}
\boldsymbol{\mathcal{C}}(\boldsymbol{x},\boldsymbol{x}',\tau)=E[\boldsymbol{q}(\boldsymbol{x},t) \, \boldsymbol{q}^*(\boldsymbol{x}',t')],
\end{equation}
where we used equation~\eqref{eq:cov-indep}, under the assumption of wide-sense stationarity of the stochastic process $\boldsymbol{q}(\boldsymbol{x},t)$ and $\tau = t - t'$ is the difference between the two times $t$ and $t'$. This implies that the covariance depends only on the difference between two times, $t$ and $t'$ and, therefore, we can write, $\boldsymbol{\mathcal{C}}(\boldsymbol{x}, \boldsymbol{x}', t, t') \rightarrow \boldsymbol{\mathcal{C}}(\boldsymbol{x}, \boldsymbol{x}', \tau)$. The last step allows  reformulating the problem in the spectral (also referred to as frequency) domain, as per equation~\eqref{eq:spod}. This significant result was first presented in ~\cite{lumley1967,lumley2007stochastic} and revisited in~\cite{towne2018spectral}. It provides eigenmodes at each frequency that inherit the same properties of the more traditional (non-spectral) POD.

The SPOD formulation just summarized leads to monochromatic SPOD modes that optimally characterize the second-order space-time moments of the continuous-time stochastic process considered~\cite{towne2018spectral}.

\subsection{Practical implementation}\label{sec:discrete}

In practice, the continuous-time stochastic process $\boldsymbol{q}(\boldsymbol{x}, t)$ introduced in section~\ref{sec:data_suitability} and used in section~\ref{sec:theory} is provided as discrete data. The discrete data consist of snapshots of the wide-sense stationary time series, $\mathbf{q}(\mathbf{x}, t_{i})$, $t_{i} = 1, \dots, N_{t}$, from which we subtract the temporal mean, $\bar{\mathbf{q}}$. Each snapshot $\mathbf{q}(\mathbf{x}, t_{i})$ is sampled at $M_{\mathrm{space}}$ spatial points with coordinates $\mathbf{x} \in \mathbb{R}^{M_\mathrm{space} \times d}$ (usually two- or three-dimensional, that is, $d=2$ or 3, respectively), and records $M_{\mathrm{vars}}$ variables. 

To derive the SPOD algorithm, we recast each snapshot of the discrete multidimensional data into a vector of dimension $\mathbf{q}(\mathbf{x}, t_i) = \mathbf{q}_{i} \in \mathbb{R}^M$, where $M = M_{\mathrm{space}} M_{\mathrm{vars}}$. We can then assemble the data matrix (also referred to as snapshot matrix),
\begin{equation}
\mathbf{Q} = [\mathbf{q}_{1}, \mathbf{q}_{2}, \dots, \mathbf{q}_{N_{t}}] \in \mathbb{R}^{M\times N_{t}}.
\label{eq:data-matrix}
\end{equation}
The data described by equation~\eqref{eq:data-matrix} can arise from simulations and observations of a wide-sense stationary stochastic system. We assumed the data to be composed exclusively of real numbers. However, this assumption is not strictly necessary, as the data can, in principle, be complex.

The first step to obtaining the discrete analog of the frequency-domain eigenvalue problem in equation~\eqref{eq:spod} consists of segmenting the data along the time direction into $L$ (possibly overlapping) blocks  
\begin{equation}\label{eq:blocks}
\mathbf{Q}^{(\ell)} = [\mathbf{q}_1^{(\ell)}, \dots, \mathbf{q}_{N_{f}}^{(\ell)}] \in \mathbb{R}^{M\times N_{f}}, \;\;\; \ell = 1, \dots, L \; \Longrightarrow
\left\{
\begin{aligned}
&\mathbf{Q}^{(1)} = [\mathbf{q}_1^{(1)}, \dots, \mathbf{q}_{N_{f}}^{(1)}] \in \mathbb{R}^{M\times N_{f}}, \\[0.5em]
& \mathbf{Q}^{(2)} = [\mathbf{q}_1^{(2)}, \dots, \mathbf{q}_{N_{f}}^{(2)}] \in \mathbb{R}^{M\times N_{f}}, \\[0.1em]
& \dots \\[0.5em]
& \mathbf{Q}^{(L)} = [\mathbf{q}_1^{(L)}, \dots, \mathbf{q}_{N_{f}}^{(L)}] \in \mathbb{R}^{M\times N_{f}}.
\end{aligned}
\right.
\end{equation}
Each data block $\ell$ in equation~\eqref{eq:blocks} contains $N_{f}$ time snapshots, overlaps by $N_{\mathrm{overlap}}$ time snapshots with the adjacent block, and is regarded as equally representative of the whole data by the ergodic assumption. Indeed, it is a realization of the stochastic process described by the discrete data $\mathbf{Q}$ in equation~\eqref{eq:data-matrix}. This approach of partitioning the time series into overlapping data blocks is the well-known Welch periodogram method~\cite{welch1967use,schmidt2020guide}. \textcolor{black}{A 50\% overlap, i.e., $N_\mathrm{overlap}=N_f/2$, is accepted best practice.} In the following, we will use the terms \textit{realization} and \textit{data block} interchangeably.  

The second step consists of applying the discrete Fourier transform (DFT) in time to each data block or realization in equation~\eqref{eq:blocks}:
\begin{equation}\label{eq:blocks-ft}
\mathbf{Q}^{(\ell)}\underbrace{\longrightarrow}_{\mathrm{DFT}}\hat{\mathbf{Q}}^{(\ell)} = [\hat{\mathbf{q}}_1^{(\ell)}, \hat{\mathbf{q}}_2^{(\ell)}, \dots, \hat{\mathbf{q}}_{N_{f}}^{(\ell)}] \in \mathbb{C}^{M \times N_{f}}, \;\;\; \ell = 1, \dots, L.
\end{equation}
We note that each  Fourier-transformed data block $\hat{\mathbf{Q}}^{(\ell)}$ contains $N_{f}$ frequencies. Wide-sense stationarity and ergodicity allow us to reorganize the Fourier-transformed data into $N_{f}$ data matrices, one per frequency, $f_k$. In particular, we collect all realizations of the DFT at the $k$-th frequency into
\begin{equation}\label{eq:blocks-k}
\hat{\mathbf{Q}}_k = [\hat{\mathbf{q}}_k^{(1)}, \hat{\mathbf{q}}_k^{(2)}, \dots, \hat{\mathbf{q}}_k^{(L)}] \in \mathbb{C}^{M \times L}, \;\;\; \text{for all frequencies $f_k$, $k = 1, \dots, N_{f}$}.
\end{equation}
For fluid mechanical and geophysical applications, we usually have $L\ll M$. For the parallelization of the SPOD algorithm we will make use of this notion.  

The third step is to construct the cross-spectral density matrix for each frequency, $f_k$. This step  corresponds to the discrete counterpart of equation~\eqref{eq:csd}, and can be readily achieved by calculating
\begin{equation}\label{eq:csd-discrete}
\hat{\mathbf{C}}_{k} = \frac{1}{L-1}\hat{\mathbf{Q}}_{k}\hat{\mathbf{Q}}^{*}_{k} \in \mathbb{C}^{M \times M}, \;\;\; \text{for all frequencies $f_k$, $k = 1, \dots, N_{f}$},
\end{equation}
where $L-1$ is a normalization factor \textcolor{black}{known as Bessel's correction}, that is only appropriate if the data is centered about the sample mean rather than the long-time (i.e., true) mean. The construction of the cross-spectral density matrix in equation~\eqref{eq:csd-discrete} finally allows us to write the discrete analog of the frequency-domain eigenvalue problem defined in equation~\eqref{eq:spod}: 
\begin{equation}\label{eq:spod-discrete}
\begin{aligned}
& \hat{\mathbf{C}}_{k} \mathbf{W} \pmb{\Phi}_{k} = \pmb{\Phi}_{k}\pmb{\Lambda}_k, \quad \text{with}\\[0.5em]
& \pmb{\Phi}_k = [\pmb{\upphi}^{(1)}_{k},\pmb{\upphi}^{(2)}_{k},\dots,\pmb{\upphi}^{(L)}_{k}] \in \mathbb{C}^{M \times L} & \text{(SPOD modes)}, \\[0.2em]
& \pmb{\Lambda}_k = \mathrm{diag}(\lambda^{(1)}_{k},\lambda^{(2)}_{k}\cdots,\lambda^{(L)}_{k}) \in \mathbb{R}^{L \times L} & \text{(modal energies)}.
\end{aligned}
\end{equation}
The SPOD modes, $\pmb{\Phi}_k$, and associated modal energies (or eigenvalues), $\pmb{\Lambda}_k$, can be computed by solving equation~\eqref{eq:spod-discrete} for each frequency, $f_k$. In practice, to alleviate the computational burden of this step, one usually turns to the method of snapshots~\cite{sirovich1987turbulence}:
\begin{equation}\label{eq:spod-discrete-snapshots}
\hat{\mathbf{Q}}^{*}_{k} \mathbf{W} \hat{\mathbf{Q}}_{k} \pmb{\Psi}_{k} = \pmb{\Psi}_{k}\pmb{\Lambda}_k, \quad \textcolor{black}{\pmb{\Phi}_k = \hat{\mathbf{Q}}_k\pmb{\Psi}_k\pmb{\Lambda}^{-1/2}_k}.
\end{equation}
By construction, for a given frequency $f_k$, the modes $\pmb{\upphi}_k$ are orthonormal, \ie $\pmb{\Phi}_{k}^{*}\mathbf{W}\pmb{\Phi}_{k} = \mathbf{I}$, where $\mathbf{I}$ is the identity matrix. Modes at different frequencies are instead not orthonormal, that is, $\pmb{\Phi}_{k_{1}}^{*}\mathbf{W}\pmb{\Phi}_{k_{2}} \ne \mathbf{I}$, where $k_{1} \ne k_{2}$, but orthonormal under the full space-time inner product.

Finally, the SPOD modes can be grouped per frequency, as follows:
\begin{equation}\label{eq:discrete_modes}
\pmb{\Phi} = [\pmb{\upphi}_{1}, \dots, \pmb{\upphi}_{N_{\text{f/2}}}]  
= [
\underbrace{\upphi^{(1)}_{1}, \dots, \upphi^{(L)}_{1}}_{\pmb{\upphi}_{1}}, 
\underbrace{\upphi^{(1)}_{2}, \dots, \upphi^{(L)}_{2}}_{\pmb{\upphi}_{2}}, \dots, 
\underbrace{\upphi^{(1)}_{N_{\text{f/2}}}, \dots, \upphi^{(L)}_{N_{\text{f/2}}}}_{\pmb{\upphi}_{N_{\text{f/2}}}}],
\vspace{-0.3cm}
\end{equation}
where, for real data, $\mathbf{Q}\in\mathbb{R}^{M\times N_{t}}$, the total number of frequencies is $N_{f/2} = \lceil \frac{N_{f}}{2} \rceil+1$. This is because the transformed data at negative frequencies correspond to the conjugates of the positive frequencies, and it is therefore redundant. For additional details on the SPOD method, the interested reader can refer to~\cite{towne2018spectral,schmidt2020guide}.

\subsection{SPOD for data compression}

As shown in~\cite{nekkanti2021frequency}, and further explored in~\cite{lario2022neural}, it is possible to compute a matrix of expansion coefficients $\mathbf{A}$. This can be constructed using a weighted oblique projection of the data onto the modal basis
\begin{equation}\label{eq:coeffs}
    \mathbf{A} = (\boldsymbol{\Phi}^{*}\mathbf{W}\boldsymbol{\Phi})^{-1}\boldsymbol{\Phi}^{*}\mathbf{W}\mathbf{Q} 
    = [
    \underbrace{a^{(1)}_{1}, a^{(2)}_{1}, \dots, a^{(L)}_{1}}_{\mathbf{a}_{1}}, 
    \underbrace{a^{(1)}_{2}, a^{(2)}_{2}, \dots, a^{(L)}_{2}}_{\mathbf{a}_{2}}, \dots, 
    \underbrace{a^{(1)}_{N_{f}}, a^{(2)}_{N_{f}}, 
    \dots, a^{(L)}_{N_{f}}}_{\mathbf{a}_{N_{f}}}].
\vspace{-0.3cm}
\end{equation}
where $\mathbf{A} \in \mathbb{C}^{(L\times N_{f})\times N_t}$ is the matrix containing the expansion coefficients and ${\boldsymbol{\Phi}} \in \mathbb{C}^{M \times (L\times N_{f})}$ is a matrix which gathers all the SPOD modes arranged by frequency as in equation~\eqref{eq:discrete_modes}. The full matrix of expansion coefficients, constructed using all modes and frequencies, has dimensions $L\times N_{ {f}} \times N_t$. In practice, it is common to use only a portion $L_{r}$ of the total number of modes, and (eventually) a portion $N_{f_{r}}$ of the total number of frequencies, where we denote with $\boldsymbol{\Phi}_r$ and $\mathbf{A}_r$, the reduced number of SPOD modes and expansion coefficients, respectively. This reduction recasts the original high-dimensional data into a smaller SPOD latent space of dimension $L_{r} \times N_{f_{r}}$. 

Once both SPOD modes and expansion coefficients are available, it is possible to reconstruct the original high-dimensional data as follows 
\begin{equation}
\tilde{\mathbf{Q}} = \boldsymbol{\Phi}_r {\mathbf{A}}_r,
\label{eq:SPODrec}
\end{equation}
where $\tilde{\mathbf{Q}}$ is an approximation of the original data $\mathbf{Q}$, given the truncation imposed on the number of SPOD modes and frequencies.  
The storage of the expansion coefficients and SPOD modes required to reconstruct the high-dimensional data in equation~\eqref{eq:SPODrec}, can  potentially lead to significant savings in terms of memory storage. In addition, the ability to truncate number of SPOD modes and number of frequencies can be beneficial to just store the frequencies of interest (e.g., removing high-frequency noise from the data), and capture low-rank behaviour of the process under study~\cite{towne2018spectral,schmidt2019spectral}. 

For instance, if we consider a problem constituted by 30,000 time snapshots,  37,000,000 spatial points, and 1 variable, we have: $N_t \times M_{\mathrm{space}} \times M_{\mathrm{vars}} = 30,000 \times 37,000,000 \times 1$. To store this dataset in memory, we require approximately $8.88$ TB in double floating-point precision, and $4.44$ TB in single floating-point precision.  

However, if we, for instance, store the first 3 SPOD modes and 100 frequencies of interest, the amount of storage memory required is significantly lowered. To store the SPOD modes, we would need \textcolor{black}{$3 \times 100 \times 37,000,000 \times 2$}, that leads to \textcolor{black}{$0.18$} TB  of memory in double floating-point precision, and \textcolor{black}{$0.089$} TB in single floating-point precision. To store the time coefficients, we would need \textcolor{black}{$3 \times 100 \times 30,000 \times 2$}, that leads to \textcolor{black}{$0.00014$} TB in double floating-point precision and \textcolor{black}{$0.000072$} TB in single floating-point precision. Hence, if we store 3 SPOD modes and 100 frequencies, the storage memory required is only \textcolor{black}{2}\% of the memory required to store the original dataset, and the data compression \textcolor{black}{i}s achieved by having full control on what modes and frequencies are left out (if any).

\section{Parallelization strategy}\label{sec:parallel-spod}

The parallelization efforts have been mainly focused on the matrix operations required by the SPOD algorithm, and on the data Input/Output (I/O). We outline the parallelization of the algorithm first (section~\ref{subsec:parallel-spod}), and reserve section~\ref{subsec:parallel-io} for I/O, as it is a crucial aspect to achieve competitive scalability results.

\subsection{SPOD algorithm}\label{subsec:parallel-spod}
The parallelization strategy for the SPOD algorithm uses a single-program multiple-data (SPMD) approach (the interested reader can also refer to other parallelized modal decompositions, such as~\cite{sayadi2016parallel}, that is specialized to boundary-layer flows). It allows maintaining the structure of the code similar to that of the serial one, only introducing parallel communication and synchronization in a limited number of places. For the SPOD algorithm, this  consists of decomposing the spatial dimensions $M_{\mathrm{space}}$ (conveniently flattened) of the data matrix in equation~\eqref{eq:data-matrix}. This is a practical and advantageous choice as it allows preserving all operations in time -- in particular, the DFT -- without needing expensive all-to-all communication patterns. Additionally, the number of spatial points $M_{\mathrm{space}}$ is usually much larger than the number of variables $M_{\mathrm{vars}}$, thus allowing for considerably higher parallelism. The decomposition of the space dimension allows a straightforward parallel implementation that only requires one single MPI collective reduction operation. Most of the MPI operations are implemented in an auxiliary utility module. This allows using MPI routines off-the-shelf, requiring minor modifications to the original (serial) code. 

\begin{figure}[H]
\centering
\includegraphics[width=1.0\textwidth]{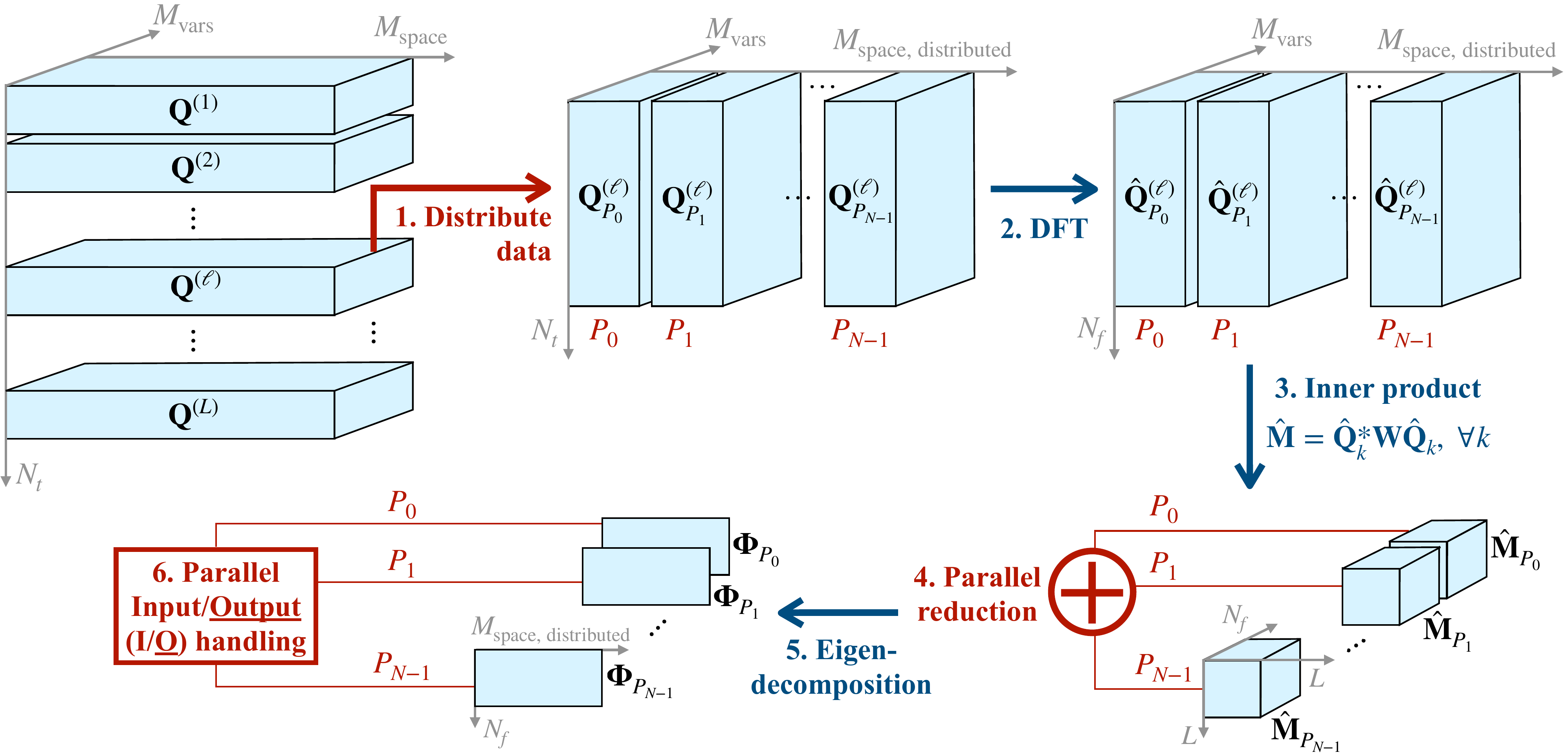}
\caption{Schematic of the parallel SPOD algorithm. The key aspect is to obtain an appropriate data decomposition layout that allows preserving all time operations as done in serial (i.e., the DFT), and decompose only the spatial dimensions of the data. Once the data is in the required parallel layout, the parallelization of the SPOD algorithm becomes trivial and consists only of a parallel reduction (step 4) of the inner product (step 3).}
\label{fig:data-layout-operations}
\end{figure}

Entering into more detail, and assuming that the data distribution is achieved right after I/O, the parallelization of the algorithm becomes trivial. Indeed, it consists of a simple parallel reduction operation for the inner product derived from the method of snapshots in equation~\eqref{eq:spod-discrete-snapshots}. The MPI-based implementation uses the mpi4py package~\cite{mpi4py,mpi4py-12years}. In figure~\ref{fig:data-layout-operations}, we depict the parallel SPOD  algorithm. In red, we represent the parallel operations required, while in blue are operations that remain unchanged compared to the serial code. The six steps reported in figure~\ref{fig:data-layout-operations} are described in the following.
\begin{itemize}
\item Step 1) Distribute data: it consists of distributing the spatial dimension of each of the data blocks $\mathbf{Q}^{\ell},\; \ell = 1, \dots, L$  across the MPI ranks available. 
\item Step 2) DFT (see equation~\eqref{eq:blocks-ft}): it consists of performing the DFT along the time dimension. This operation remains unchanged as time has not been distributed. 
\item Steps 3, 4) Inner product and reduction (see \textcolor{black}{equation~\eqref{eq:spod-discrete-snapshots}}): the inner product operation involves contracting the spatial dimension, hence it requires a parallel reduction operation.
\item Step 5) Eigen-decomposition (see equation~\eqref{eq:spod-discrete-snapshots}): this operation does not require any parallel handling as there is no manipulation of the distributed spatial dimension. Since the size of the blocks is typically small, eigen-decompositions are cheap to compute. Therefore, we perform this step redundantly in all MPI ranks. If this task ever becomes a computational bottleneck, it would be straightforward to replace with a \textcolor{black}{distributed-memory}  implementation exploiting the \textcolor{black}{already} available \textcolor{black}{CPU resources, or alternatively, GPU off-loading for acceleration}.
\item Step 6) Parallel I/\underline{O}: This operation involves writing the SPOD modes to disk. The I/O handling is described in the next section (\ref{subsec:parallel-io}), and is key to the scalability and overall performance of our implementation.
\end{itemize}

\subsection{I/O handling}\label{subsec:parallel-io}
We dedicate a separate section to the I/O aspect of PySPOD as, in most practical scenarios, it plays a critical role in the performance of PySPOD overall. This is especially prominent in a strong scaling scenario such as that presented in section~\ref{sec:strong}. We divide this section into background and related work (section~\ref{sec:related}), and the PySPOD I/O implementation (section~\ref{sec:pyspod-io}).

\subsubsection{Background and related work} \label{sec:related}
Since the early 1990s, I/O has been acknowledged as a significant contributor to parallel application performance \cite{io_characteristics_1995}. However, relatively little attention was given to I/O compared to the compute and communication subsystems during that time \cite{two-phase}.
Since then, the performance gap between the I/O throughput and the compute capability of supercomputers has only increased, hence exacerbating data movement challenges \cite{adios}. 
A 2012 report titled ``Storage challenges at Los Alamos National Lab'' noted that parallel file systems lack large parallel I/O, which would be both high bandwidth and resistant to I/O patterns without additional tuning \cite{challenges_lanl}.
Consequently, I/O systems remain relatively slow and complex, making it difficult for non-expert users to utilize them fully.

Middleware libraries were introduced to improve usability and simplify parallel I/O. After all, as \cite{adios} noted, users need not be aware of their data's low-level layout and organization, which opens possibilities for specialized I/O middleware.
MPI-IO enables simple N--1 I/O ($N$~processes, $1$~file)~\cite{thakur1999implementing}, and it has been an ongoing effort since 1999. It contains optimizations reducing the number of distinct I/O requests using techniques such as data sieving and collective I/O for noncontiguous accesses \cite{thakur2002optimizing}. MPI-IO, however, is still low-level, as the developer needs to calculate the offsets within the file, while MPI aims to access the data efficiently.
Higher-level libraries such as HDF5 \cite{hdf5}, Parallel netCDF \cite{pnetcdf}, ADIOS \cite{adios}, and ADIOS~2 \cite{adios2} have also been introduced to isolate developers from low-level file system details and alleviate the burden of platform-specific performance optimization. These libraries often build on top of MPI-IO and introduce custom file formats.

Although robust solutions for parallel I/O are available, achieving good performance often requires application-specific and system-specific tuning. Researchers have dedicated significant effort to optimizing applications that utilize the popular Lustre file system.
Tools such as Lustre IO Profiler \cite{lioprof} were introduced to monitor and understand the I/O activities on the file system. The authors of this tool highlight the importance of incorporating system-specific details, such as the number of Object Storage Targets (OSTs), into the MPI distribution used in order to improve the achieved I/O bandwidth.
The authors of Y-Lib \cite{ylib} also point out the low performance of MPI-IO on Lustre file systems and propose a solution that minimizes contention for file system resources by controlling the number of OSTs with which each aggregator process communicates. This solution is shown to outperform MPI-IO in many cases.
Similarly, the authors of \cite{balle_wrf} demonstrate that Parallel netCDF with MPI-IO does improve the performance over netCDF; however, even with supercomputer-specific optimizations, the performance still disappoints once a large number of MPI ranks is used. The authors utilize asynchronous I/O (quilt) servers alongside Parallel netCDF to further improve the effective bandwidth.
Other solutions explore tuning the middleware or file system parameters, such as those presented in \cite{io_tuning_cray,tuning_hdf5_lustre,taming_tuning,autotuning}. These approaches often use sophisticated techniques, such as genetic algorithms, to identify and tune parameters of the I/O stack (HDF5, MPI-IO, and Lustre/GPFS parameters) \cite{taming_tuning,autotuning}. 
The wealth of work on the topic highlights the difficulty of achieving good performance on current parallel file systems.

Despite the availability of powerful solutions and tools to tune the performance, many developers still choose not to use them, even in leadership scale systems \cite{lioprof}. An analysis conducted in 2015 using Darshan \cite{gropp_darshan} found that, depending on the supercomputing facility, between 50 to 95\% of monitored jobs used POSIX I/O exclusively.
%
%
%
However, as \cite{gropp_darshan} notes, relying on POSIX I/O does have to result in poor application performance. In fact, in some cases, using a naive 
N--N ($N$ processes, $N$ files) approach, where each process reads from or writes to a unique file, can achieve better performance and scalability than N--1, as it does not incur overhead to maintain data consistency. This has been observed in PanFS, GPFS, and Lustre file systems. However, the N--N approach presents challenges related to usability, such as when restarting with a different number of processes, and high metadata costs at scale \cite{challenges_lanl,cosmology_io}.

\subsubsection{PySPOD I/O}\label{sec:pyspod-io}
In PySPOD, we take advantage of the N--N model's good performance, but like authors of \cite{cosmology_io}, slightly modify it to be N--M (generally M $<$ N) mapping, where M is not to be confused with `italic' $M$ adopted in section~\ref{sec:discrete}. Such a setting, where N is the number of processes and M is the number of files, is better suited to our application characteristics.
We use a simple two-phase I/O remnant of that proposed in \cite{two-phase} back in 1993.

In particular, we first read the data from disk in a contiguous manner; afterwards, we re-distribute the data according to the parallel decomposition our application needs, as explained in section~\ref{subsec:parallel-spod}. Parallel data re-distribution can be performed either with collective communications, or point-to-point communications. In our case, we use non-blocking point-to-point MPI communications, as it results in higher reliability in the HPC systems that have been used in this work.

In the following, we enumerate the reasons for our decision to adopt this two-phase I/O, and categorize these reasons into \textit{application-specific} and \textit{performance-specific}.
\begin{description}
\item \textit{Application-specific}
\begin{itemize}
    \item PySPOD is designed to analyze large datasets, which are often split over multiple files. One such example is the Climate Data Store (CDS), which provides rich climate datasets~\cite{cds_levels}, and that has been used in the scalability analyses presented in sections~\ref{sec:strong} and \ref{sec:weak}. CDS, however, limits the amount of data that can be downloaded in a single request. The resulting datasets' size can be in the order of tens of terabytes split over several hundred files. Another example arises from computational fluid dynamics (CFD) simulators such as SSDC, Nektar++ and Charles~\cite{ssdc,moxey2020nektar++,BresEtAl2017AIAAJ}. Such software often writes output using one file per requested timestep.
    Considering the data size and PySPOD's competitive scalability performance, we decided against converting the data to a common format (such as HDF5), as the conversion could be more expensive than the analysis, and we implemented several native readers for different data formats.
    
    \item Using two-phase I/O allows reusing sequential readers. The design of keeping the I/O (first phase) separate from the data distribution (second phase) effectively makes the I/O phase into many concurrent but sequential I/O streams. This design makes it easy for a user or a developer to implement support for additional file formats in PySPOD. The programming burden is greatly reduced, as there are plenty of sequential reader modules in the Python Package Index which can be used with minimal effort. This design choice removes the burden of ``thinking parallel'' from a user/developer perspective, as the data distribution is performed in the second phase, and it does not depend on the file format.
\end{itemize}

\item \textit{Performance-specific}
\begin{itemize}
    \item Using a two-phase I/O results in fewer and more contiguous requests to the storage system, which is preferred on a parallel file system. For illustration, most often, the data is split over multiple files, where each file represents a different range of timesteps, and the spatial coordinates follow the first dimension. In our design, each process reads contiguous data from at least one input file, i.e., all spatial dimensions for a subset of timesteps (or, more generally, the first dimension, i.e., time), which is good for performance. An alternative design, such as using MPI-IO to immediately read only a subset of spatial variables for all timesteps, may result in every process accessing every file, depending on the MPI implementation used. \textcolor{black}{As we discussed in section \ref{sec:related}, MPI-IO has been shown to perform poorly on Lustre file systems~\cite{ylib}.} In fact, \textcolor{black}{an MPI-IO reader} is also implemented in PySPOD; however, \textcolor{black}{we also obtained poor performance results} in early experimentation; hence, we decided to extend and optimize the two-phase reader.

    \item Separating I/O from MPI removes the risk of performance degradation due to an underoptimized MPI distribution. As discussed in section~\ref{sec:related}, MPI-IO's performance may depend heavily on the MPI distribution used and its knowledge of the underlying I/O system. By separating I/O and data distribution, we read the data from the disk efficiently and then explicitly distribute it with MPI. Since we use point-to-point communication, which is at the very core of MPI, it is much less likely that those functions will perform poorly.
\end{itemize}

\item \textit{Application-specific} and \textit{performance-specific}
\begin{itemize}
    \item SPOD output is expected in the form of one output file per frequency and mode. 
    However, because the data for each frequency is spread across all processes, this would lead to N--1 storage access for each file. This approach would be highly inefficient due to relatively small spatial dimensions. To address this issue, we use a two-phase process similar to the one used for reading files, allowing each file to be written by only one process.
\end{itemize} 
\end{description}

To address the memory capacity limitations of PySPOD, we implemented several precautions. One such measure was to read data in chunks, with each process reading approximately 256~MB of data (value determined empirically). This helps reducing memory overhead since processing each chunk requires twice the memory, as time-split data being sent can only be deallocated once it is received by the target process, and space-split data is being received simultaneously. \textcolor{black}{The chunk size should be as small as possible to reduce the memory overhead of using auxiliary memory but large enough to hide the I/O latency. On our system, 256~MB chunks correspond to roughly 6\% of total memory per CPU core.}

Additionally, we used a dictionary of NumPy arrays instead of one large array. Such design allows us to deallocate processed data as its FFT transforms replace it in memory, therefore reducing peak memory usage. Specifically, we store the data for each chunk in three dictionary keys, with each NumPy array occupying approximately 85~MB of memory.

Moreover, we implemented reading in such a way that one process first seeks through the metadata to identify which range of timesteps is available in which file. This information, broadcast with MPI, significantly reduces the overhead associated with N processes opening all M files.

\section{Datasets} \label{sec:datasets}
The datasets adopted to test the parallel SPOD algorithm consists of fluid mechanics, and geophysical data. The former uses jet data produced by high-fidelity large-eddy simulation (LES), and is described in section~\ref{sec:fluidmechanics}. The latter uses fifth-generation reanalysis data (ERA5) produced by ECMWF~\cite{hersbach2020era5}, and is described in section~\ref{sec:geophysical}, along with associated SPOD results.

\subsection{SPOD analysis of fluid mechanics datasets}\label{sec:fluidmechanics}
In our first example, we analyze the data generated by Yeung et al.~\cite{YeungEtAl2022AIAA} using the solver Charles \cite{BresEtAl2017AIAAJ}, from LES of a supersonic twin-rectangular jet at a Reynolds number of $Re\approx10^6$ based on the jet exit conditions and the equivalent nozzle diameter. The simulation was previously validated by Br\`es et al.~\cite{BresEtAl2021AIAA} against the companion experiments of Samimy et al.~\cite{SamimyEtAl2023JFM}. The time-resolved data consist of 20,000 snapshots of the 3D flow field, interpolated onto a Cartesian grid and saved at a time interval of $\Delta t=0.2$. Each snapshot records five primitive variables: density ($\rho$), velocities ($u,v,w$), and temperature ($T$), in single precision, for a total of $N_x\times N_y\times N_z\times N_\mathrm{var}=625\times270\times344\times5$ data points per snapshot. Storage of the database requires 18,392 GB on disk in HDF5 format, or \textcolor{black}{23,220} GB once loaded into memory.
\begin{figure}[H]
\centering
\includegraphics{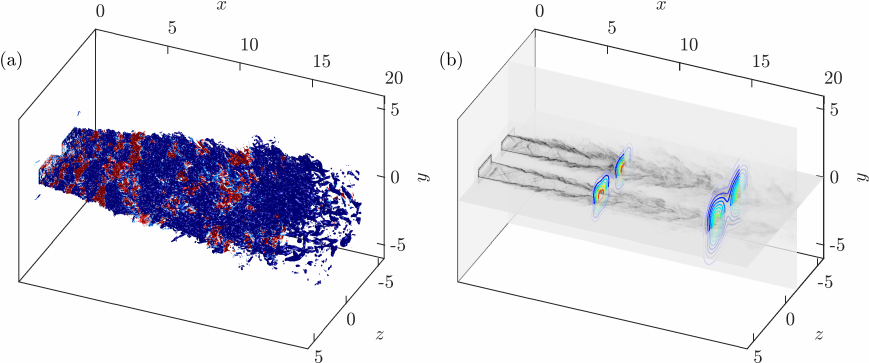}
\caption{Instantaneous flow field of the twin-rectangular jet: (a) Q-criterion isosurface, colored by pressure; (b) numerical schlieren on the $y=0$ and $z=-1.8$ planes, with contours of mean streamwise velocity on the $x\in\{ 8,16 \}$ planes.}
\label{fig:jetQcritSchlieren}
\end{figure}
Figure~\ref{fig:jetQcritSchlieren}(a) visualizes an instantaneous Q-criterion isosurface, showing the highly turbulent flow field of the high-Reynolds number jet. The colors represent pressure, $p-p_\infty$. Alternating bands of red and blue in the region $x\lesssim5$ correspond to near-field coherent pressure fluctuations, which radiate sound outward and contribute to far-field noise. Figure~\ref{fig:jetQcritSchlieren}(b) shows planar slices of the instantaneous density gradient magnitude, $|\nabla\rho|$, i.e., an artificial schlieren. Shock cells can be observed inside the potential cores. Superimposed on the schlieren are contours of the mean streamwise velocity, $\bar{u}$. Despite the chaotic instantaneous flow field, $\bar{u}$ displays reflectional symmetries about the major and minor axes, $y=0$ and $z=0$, respectively. The mean flow thus recovers the geometrical symmetries of the twin-rectangular nozzles.

The modal decomposition of axisymmetric jets is typically preceded by an azimuthal Fourier transform, which exploits the rotational invariance of the turbulent statistics. Without loss of generality, the transform reduces the analysis from a single 3D SPOD to one 2D SPOD per azimuthal wavenumber. The absence of azimuthal homogeneity in the twin-rectangular jet precludes such a simplification, thus necessitating a costly 3D analysis, to which PySPOD is ideally suited. To perform SPOD, we assemble the primitive variables into the state vector $\vb{q} = \mqty[\rho,u,v,w,T]^\mathrm{T}$.
Since the flow is compressible, we choose the weight matrix
\begin{equation}
    \vb{W} =
\int_z\!\int_y\!\int_x \mathrm{diag}\qty( \frac{\overline{T}}{\gamma\overline{\rho}M_j^2}, \overline{\rho}, \overline{\rho}, \overline{\rho}, \frac{\overline{\rho}}{\gamma\qty(\gamma-1)\overline{T}M_j^2} ) \dd{x}\dd{y}\dd{z}, \label{eq:chuNorm}
\end{equation}
such that the inner product $\expval{\vb{q}_1,\vb{q}_2}=\vb{q}_1^\mathrm{H}\vb{W}\vb{q}_2$ induces the compressible energy norm \cite{Chu1965ActaMech}. Here, $\gamma=1.4$ is the adiabatic index; $M_j=1.5$ is the jet Mach number. We select a block size of $N_f=256$, with 50\% overlap, giving $L=155$ blocks.

\begin{figure}
\centering
\includegraphics{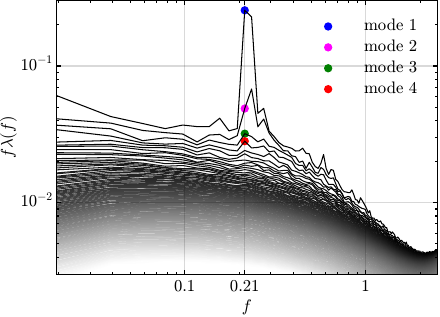}
\caption{\textcolor{black}{Premultiplied SPOD eigenvalue spectra of the twin-rectangular jet. The spectra fade from black to white with increasing mode number. Modes corresponding to the highlighted ($\bullet$) eigenvalues at $f=0.21$ are reported in figure~\ref{fig:jetF0011Mode}.}}
\label{fig:jetSpec}
\end{figure}

The premultiplied SPOD eigenvalue spectra are reported in figure~\ref{fig:jetSpec}. The leading eigenvalues show a prominent peak at $f\approx0.2$, where $f$ is nondimensionalized by the nozzle height and the ambient speed of sound. This peak corresponds to the fundamental screech tone. Screeching of the supersonic twin-rectangular jet stems from acoustic resonance between the nozzle and the shock cells, and has been observed both experimentally \cite{JeunEtAl2022AIAAJ,SamimyEtAl2023JFM} and numerically \cite{BresEtAl2021AIAA,BresEtAl2022AIAA,YeungEtAl2022AIAA} to occur at a similar frequency. 
The peak at $f\approx0.2$ persists at least to the second eigenvalue. Large separations between the first, second, and third eigenvalues---termed low-rank behavior \cite{SchmidtEtAl2018JFM}---in the frequency range $0.2\lesssim f\lesssim0.3$ signal the presence of coherent structures arising from underlying physical instabilities, in this case the screech mechanism.

\begin{figure}
\centering
\includegraphics{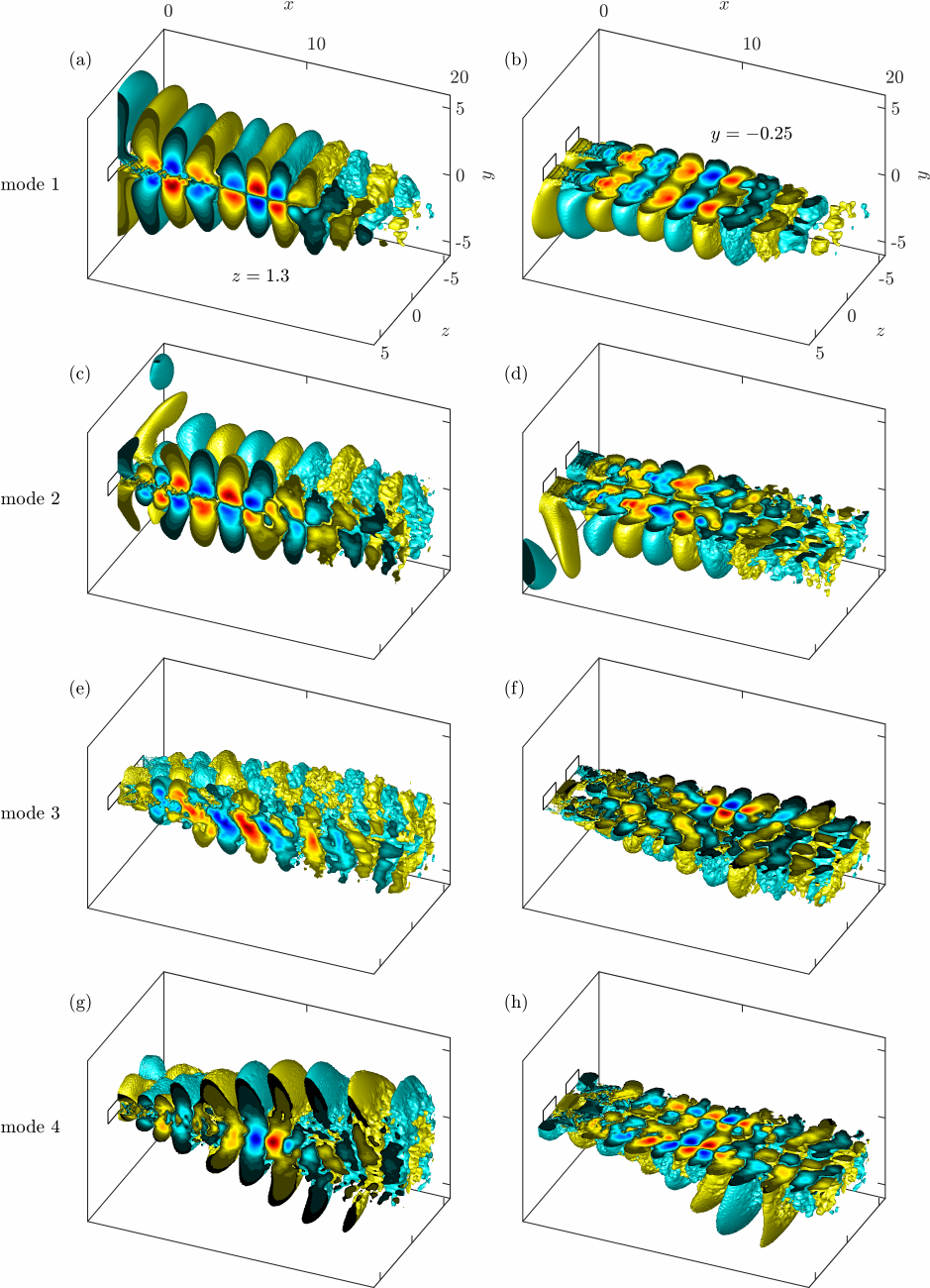}
\caption{SPOD modes of the twin-rectangular jet at frequency $f=0.21$: (a,b) mode 1; (c,d) mode 2; (e,f) mode 3; (g,h) mode 4. Isosurfaces of $\mathrm{Re}\{\vb*{\phi}_p\}=\pm 0.0005$ are shown, along with cross-sections at $z=1.3$ (left column) and $y=-0.25$ (right column). The corresponding SPOD eigenvalues are highlighted in figure~\ref{fig:jetSpec}.}
\label{fig:jetF0011Mode}
\end{figure}

In figure~\ref{fig:jetF0011Mode}, we study these structures by visualizing the pressure component, $\vb*{\phi}_p$, of the SPOD modes corresponding to the first four eigenvalues at $f=0.21$. We recover $\vb*{\phi}_p$ from the density and temperature components of each mode, $\vb*{\phi}_\rho$ and $\vb*{\phi}_T$, respectively, using the linearized ideal gas equation,
\begin{equation}
    \vb*{\phi}_p = \frac{1}{\gamma}\qty(\vb*{\phi}_\rho \overline{T} + \overline{\rho} \vb*{\phi}_T)\,.
\end{equation}
Isovalues of $\mathrm{Re}\{\vb*{\phi}_p\}=\pm 0.0005$ are chosen to highlight the 3D structure of the far-field acoustic waves. Each row in figure~\ref{fig:jetF0011Mode} shows one mode. The left and right columns, \ref{fig:jetF0011Mode}(a,c,e,g) and \ref{fig:jetF0011Mode}(b,d,f,h), also display the cross-sectional views of the planes $z=1.3$ and $y=-0.25$, respectively. These provide insights into the wavepackets within the jet plume, which are well-known to be efficient sources of noise \cite{JordanColonius2013AnnuRev}. In figure~\ref{fig:jetF0011Mode}(a,b), mode 1 recovers near-perfect antisymmetry about the major-axis plane, $y=0$, and symmetry about the minor-axis plane, $z=0$. In contrast, mode 2 in \ref{fig:jetF0011Mode}(c,d) is clearly antisymmetric about both planes. Mode 4 in \ref{fig:jetF0011Mode}(g,h), on the other hand, is symmetric about both planes. The symmetry of mode 3 is difficult to ascertain visually from \ref{fig:jetF0011Mode}(e,f). In particular, the strong in-phase (\ref{fig:jetF0011Mode}(b,h)) and out-of-phase (\ref{fig:jetF0011Mode}(d)) coupling between the twin jets observed in modes 1, 2, and 4 appear to be lost in mode 3 (\ref{fig:jetF0011Mode}(f)). This is due to insufficient statistical convergence.

\textcolor{black}{A comprehensive discussion of twin-rectangular jet dynamics exceeds the scope of this work. However, our findings illustrate some of the physical insights that can be gleaned from a 3D modal analysis, which PySPOD vastly accelerates.}

\subsection{SPOD analysis of geophysical datasets}\label{sec:geophysical}

In our second example, we use data from ERA5, a fifth-generation reanalysis dataset produced by the ECMWF that combines model data with global observations using the laws of physics to create a complete and consistent global climate and weather dataset. We obtained this data from the Climate Data Store (CDS)~\cite{cds_levels}.

\begin{figure}[H]
\centering
\includegraphics[width=0.49 \linewidth]{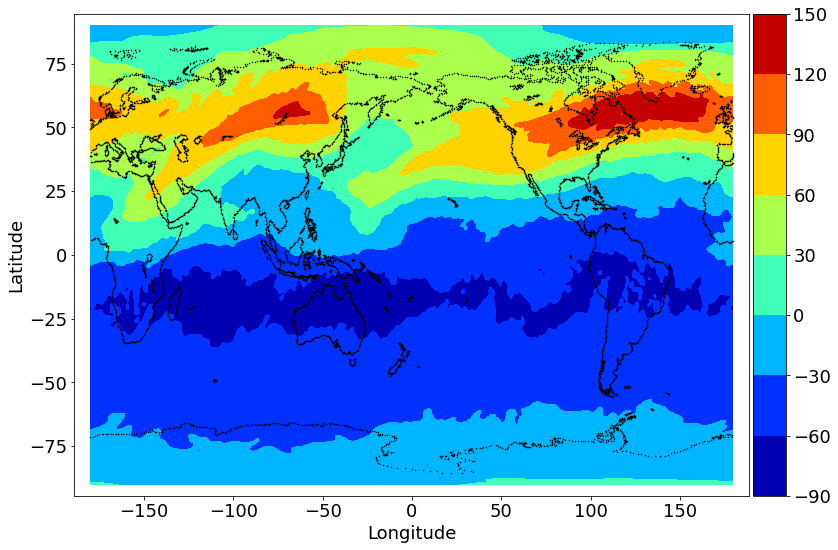}
\includegraphics[width=0.49 \linewidth]{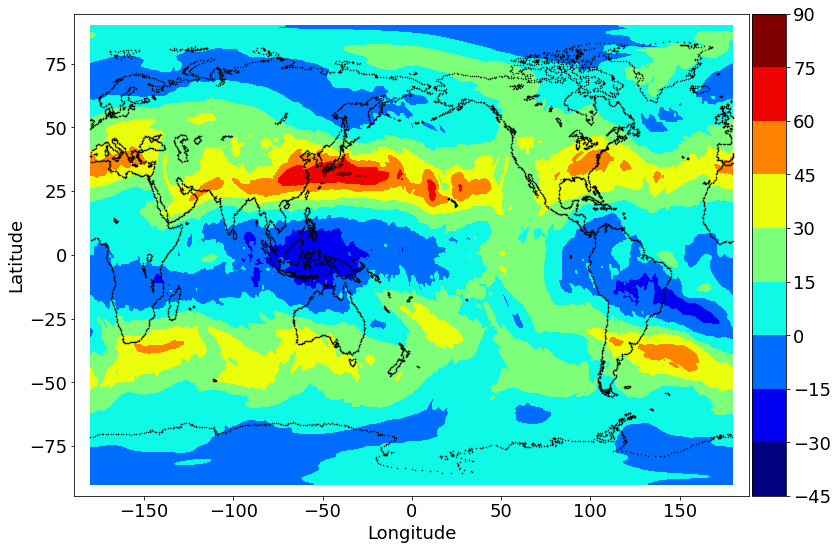}
\includegraphics[width=0.49 \linewidth]{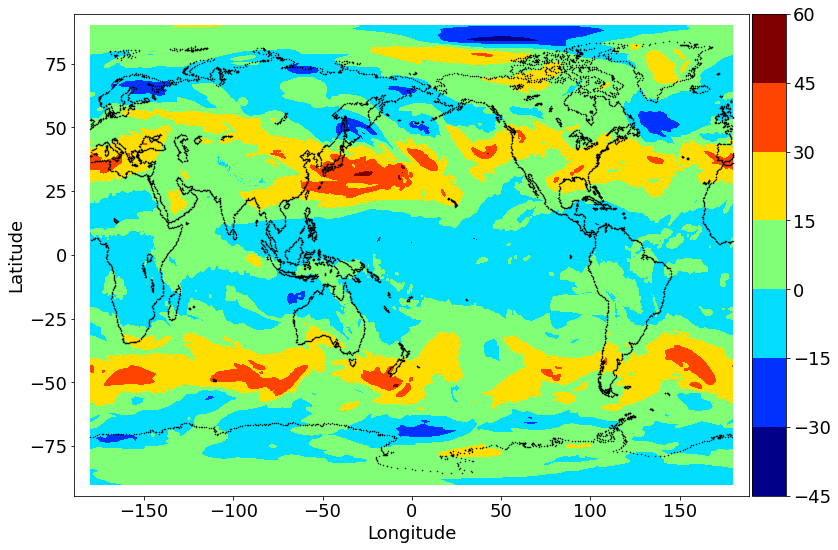}
\includegraphics[width=0.49 \linewidth]{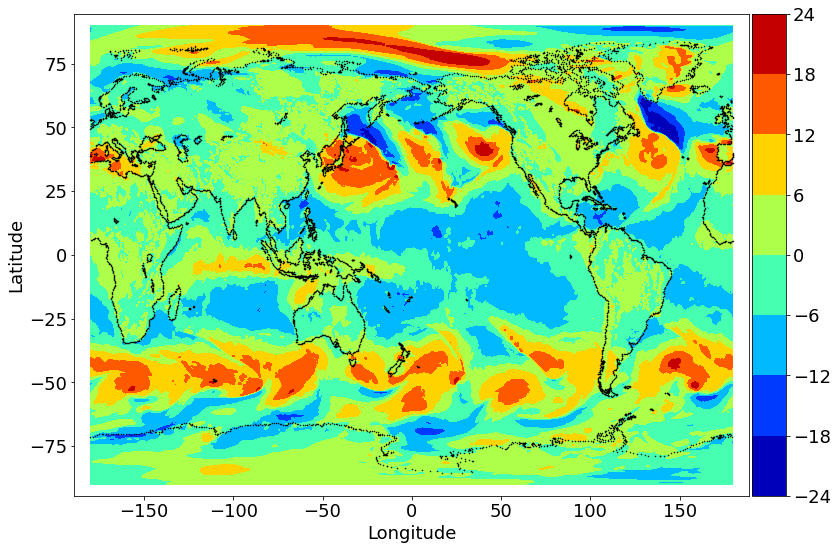}
\caption{U component of the wind velocity for pressure (i.e., vertical) levels 1 (top left), 12 (top right), 24 (bottom left), and 37 (bottom right), at midnight (00:00) on the 1st of January 2010. Level 1 corresponds to a pressure of 1 millibars, level 12 to 125 millibars, level 24 to 600 millibars, and level 37 to 1000 millibars.}
\label{fig:geo-snapshots}
\end{figure}

In particular, we consider the horizontal speed of air moving towards the east on 37 pressure levels (i.e., vertical levels) spanning from January 1940 to December 2022 \cite{cds_levels}. This quantity is also referred to as U component of the horizontal wind velocity, and its unit is meters per second. The dataset contains 727,584 time snapshots on a 3D grid of dimension 1440$\times$721$\times$37. The total size of this dataset is 51,745~GB, corresponding to 27.9 trillion data points and 103,490~GB in memory when using single-precision floating-point, as for this example, and 199,452~GB in double-precision. In figure~\ref{fig:geo-snapshots}, we depict the horizontal wind velocity adopted for pressure level 1 (top left), 12 (top right), 24 (bottom left), and 37 (bottom right). 
dataset.

The SPOD algorithm uses 10-year data blocks, that corresponds to a block size of $N_f = 87,600$ time snapshots, resulting in a total of $L = 8$ blocks, where we used 0\% overlapping. 

\textcolor{black}{This SPOD configuration was chosen to capture low-frequency atmospheric modes. More specifically,} the analysis of this dataset aims to capture the quasi-biennial oscillation (QBO) that has an approximate period of 2 to 2.5 years, as reported in~\cite{schmidt2019spectral}. This atmospheric oscillation is characterized by quasi-periodic reversals of the zonal-mean zonal winds in the equatorial stratosphere -- see also~\cite{baldwin2001quasi}. QBO has important implications on teleconnections, influencing weather patterns in the Northern Hemisphere, and the tropics (including tropical precipitation)~\cite{gray2018surface}. Its most distinctive feature is a latitudinal band of the U component of the wind velocity in the tropical region ($\pm 20^{\circ}$ latitude).

\begin{figure}[H]
\centering
\includegraphics[width=0.45 \linewidth]{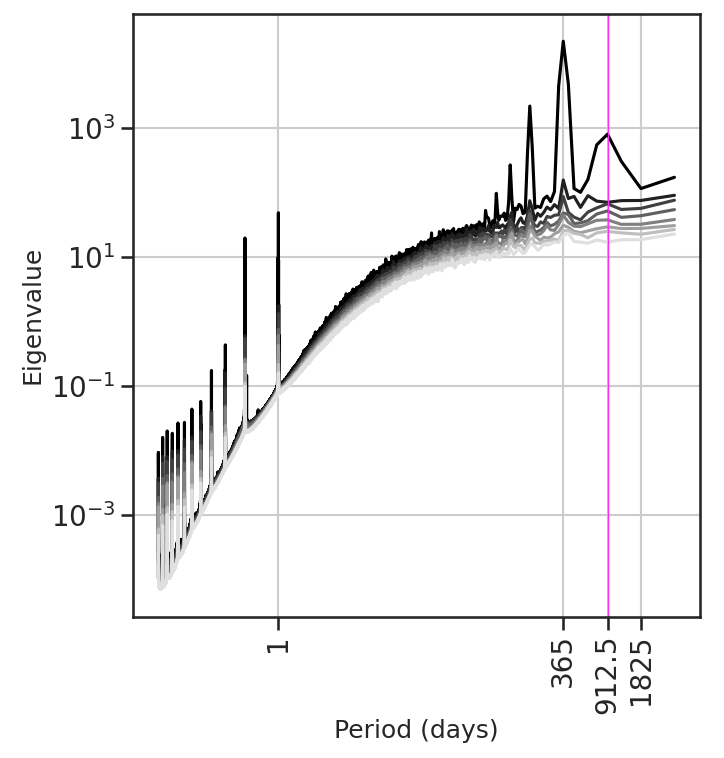}
\caption{Eigenvalue spectra vs.\ period (in days). The pink vertical line denotes the peak associated to the QBO, whose associated mode is depicted in figure~\ref{fig:qbo3d}. We can also notice other high-frequency peaks, related to yearly, sub-yearly, daily and sub-daily patterns.}
\label{fig:qbo3deigs}
\end{figure}

\begin{figure}[H]
\centering
\includegraphics[width=0.72\linewidth]{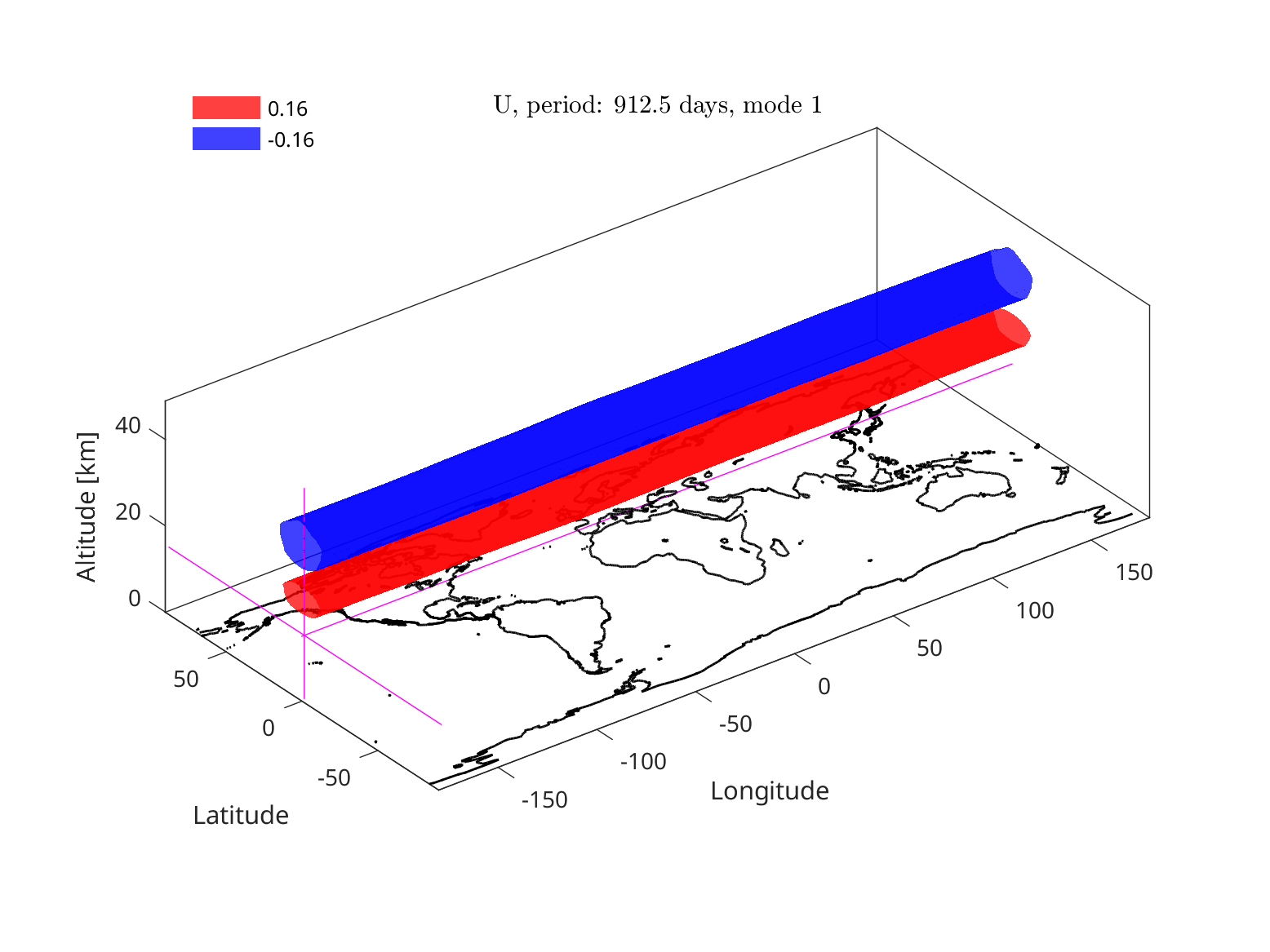}
\caption{Real part of the leading three-dimensional SPOD mode for the U component of the wind velocity; period of 912.5 days.}
\label{fig:qbo3d}
\end{figure}

Indeed, figure~\ref{fig:qbo3deigs} shows the eigenvalue spectra for the U component of the wind velocity, where the leading eigenvalue show a prominent peak at period $T = 912.5$ days. This peak corresponds to the QBO, and shows as this phenomenon exhibits low-rank behaviour, since the energy of the leading eigenvalue is remarkably separated from the other eigenvalues. Figure~\ref{fig:qbo3d} shows the highly-coherent three-dimensional structure of the leading mode. This manifests as a latitudinal band of the U component of the wind velocity in the tropical region, that is the signature of QBO. The results are consistent with those in~\cite{schmidt2019spectral}, albeit the data adopted there was ERA-20C~\cite{poli2016era}, that has significantly coarser spatial and temporal resolution, but longer coverage (from 1900 to 2010). In our results, we also observe several high-frequency peaks in the eigenvalues (periods shorter than 1 day) thanks to the improved temporal resolution of our dataset.

As for the example presented in section~\ref{sec:fluidmechanics}, a comprehensive discussion of QBO dynamics is outside the scope of this work.
Yet, we remark that the analysis outlined in this section would have been extremely challenging without the parallel (distributed) implementation presented in this work.

\section{Scalability} \label{sec:scalability}
The scalability results were obtained on the geophysical dataset introduced and analyzed in section~\ref{sec:geophysical}. Here, we discuss the scalability setup (section~\ref{sec:scalability_setup}), strong (section~\ref{sec:strong}) and weak (section~\ref{sec:weak}) scalability studies conducted using that dataset.

\subsection{Scalability setup} \label{sec:scalability_setup}
To test the scalability of the PySPOD library, we used the Shaheen~II supercomputer, a Cray~XC40 system hosted by King Abdullah University of Science and Technology (KAUST) \cite{shaheen2}. Each of Shaheen's 6,174 nodes features two Intel Haswell (Xeon E5-2698v3) CPUs with 16 cores each and 128~GB of memory. The nodes are connected via a Cray Aries interconnect with Dragonfly topology. While the nodes were allocated in exclusive mode by the scheduler, the network was shared with other users, as is typical in a production environment. To alleviate the memory capacity bottleneck and maximize the I/O bandwidth, we run PySPOD with 4~\textcolor{black}{MPI} processes per \textcolor{black}{compute} node. \textcolor{black}{We do not use multithreading}.

For storage, we utilized the Cray Sonexion\textregistered~2000 Storage System, which offers over 16~PB of usable capacity using 72 Scalable Storage Units (SSUs), 144 Object Storage Services (OSSs) and 144 Object Storage Targets (OSTs). The system features 5,988 4~TB disks. The theoretical performance of this storage exceeds 500 GB/s \cite{Hadri2017RegressionTO}, and published application benchmarks have shown that applications such as WRF can achieve a bandwidth of 25--35 GB/s, including communication, or 65 GB/s of raw I/O performance measured on aggregators when using 144 OSTs and 4 MPI ranks per node. Similar results have been obtained with NPB benchmarks \cite{george_bb}.

We stored both datasets on the Lustre storage system described in section~\ref{sec:scalability_setup}. Since many processes may need to access each file and the datasets are large, we used a stripe count of 144. However, we did not use striping for output to limit each process to only communicate with one Object Storage Target (OST) and reduce contention, given that each process writes a single file. In both cases, we used the default stripe size of 1 MB.

We report the following timings for PySPOD: time for I/O (combined reading and writing of 5 leading modes to disk), computation of the DFT (step 2 in figure~\ref{fig:data-layout-operations}), computation of the inner product and associated matrix $\hat{\mathbf{M}}$ (step 3 in figure~\ref{fig:data-layout-operations}), eigenvalue decomposition (step 5 in figure~\ref{fig:data-layout-operations}, and computation of the SPOD modes (just prior to step 6 in figure~\ref{fig:data-layout-operations}. Given the variability of performance on Dragonfly networks \cite{dragonfly-variability} and shared file systems \cite{io_variability}, we report arithmetic averages based on 5 repetitions.

\textcolor{black}{We employed PySPOD release 2.0.0}, along with the following packages: mpi4py~3.1.4, netCDF4~1.6.3, NumPy~1.24.2, SciPy~1.10.1, and Xarray~2023.2.0. Cray-provided modules for Python~3.10.1 and Cray MPICH~7.7.18.

\subsection{Strong scalability} \label{sec:strong}

To perform the strong scalability analysis, we used a 2D version of the dataset presented in section~\ref{sec:geophysical}, corresponding to the 10~hPa pressure level~\cite{cds_single}. This dataset contains 727,584 time snapshots on a 1440$\times$721 grid and is 1,407 GB in size when stored on the disk in netCDF format. The data set contains a total of 755.4 billion data points in single-precision floating point, which amounts to 5,628~GB in memory once stored in double-precision floating-point for this study.
As we scale from 256 up to 8,192 processes (64 to 2,048 nodes), this corresponds to between 21.99~GB and 0.68~GB, or 4,055 and 126 spatial points over time per process. This provides a broad range of scenarios. In all cases, all the temporal data is split into 1-year (8,760 snapshots) blocks.
\textcolor{black}{This choice of block size is different to that of section~\ref{sec:geophysical} for the sake of using consistent configurations in both strong and weak scalability analyses.}

\begin{figure}[H]
	\centering
    \includegraphics[width=0.72 \linewidth]{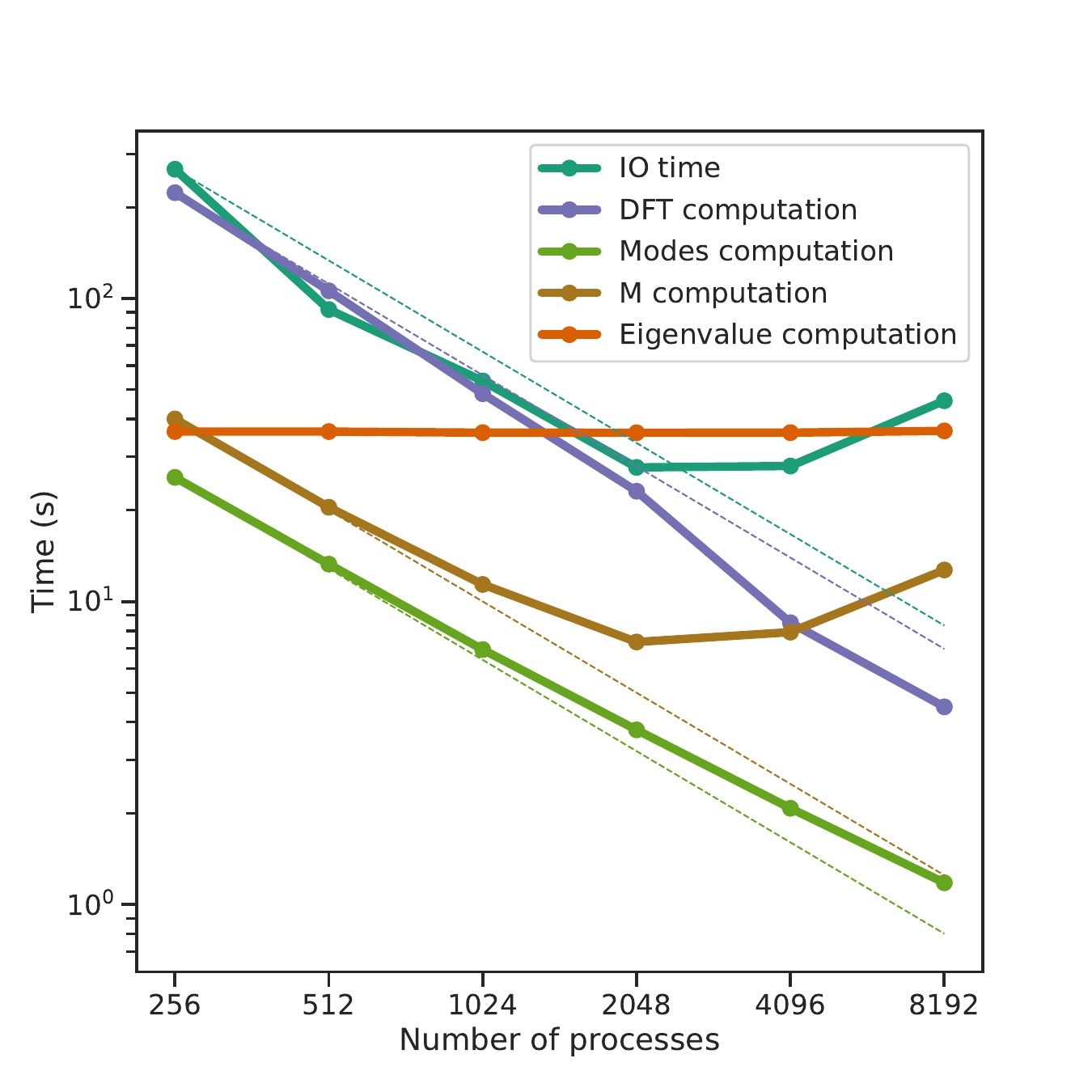}
    \caption{Strong scalability of PySPOD using horizontal speed of air moving towards the east data from January 1940 to December 2022, using from 256 to 8,192 processes. Dashed lines represent ideal scalability for each component.}
    \label{fig_strong}
\end{figure}

As shown in figure~\ref{fig_strong}, the outcomes differ depending on the task that is being performed (different curve colors in figure~\ref{fig_strong}). In particular, we achieved a satisfactory speedup for I/O as the number of processes increased (\textcolor{teal}{teal curve}), with the peak average read bandwidth of 73.76~GB/s (4,096 processes) and the maximum individual measured bandwidth of 88.04~GB/s. 
Even though the read speed peaks at 4,096 processes, the combined I/O time (reading and writing) is minimally lower when using 2,048 processes. 
The reported timings include communication (i.e., the second phase of the two-phase I/O) and should be compared to the 25--35~GB/s figure quoted in section~\ref{sec:scalability_setup}. We consider the results extremely competitive on this hardware.
%
The FFT calculation (\textcolor{violet}{violet curve}), the second most expensive part of the SPOD algorithm, exhibited surprising behavior. 
The speedup is measured at 50$\times$ when the number of processes is increased by 32$\times$ (from 256 to 8,192 processes).
This phenomenon has been previously observed on the same supercomputer \cite{mpi-fft} and can be explained by cache effects, frequency throttling, and possibly problem size-dependent NumPy optimizations.

The computation of matrix $\hat{\mathbf{M}}$ (\textcolor{brown}{brown curve}), a collective communication-heavy routine, stops scaling past 2,048 cores; however, at this point, it only takes around 7 seconds. Similarly, the eigenvalue computation (\textcolor{orange}{orange curve}), which is problem size-dependent and not parallelized (see rationale in section~\ref{subsec:parallel-spod}), is represented by a horizontal line, i.e., its cost depends on the dimensions of the data and not the number of processes.
On the other hand, the calculation of SPOD modes (\textcolor{ForestGreen}{green curve}) scaled well despite its cost being measured in single seconds.

In terms of overall relative efficiency\footnote{The relative parallel efficiency is calculated as $E_P = (256 \times T_{256}) / (P \times T_{P})$, where $T_{P}$ is the wall-clock time corresponding to $P$ processes.}, we calculate it to be 74\% when using 2,048 cores and 38\% or lower once increasing to 4,096 cores and beyond. 
Even though the efficiency is only 15\% when using 8,192 processes, we note that in this scenario, the total runtime is only around 127 seconds. The shortest runtime was achieved when using 4,096 processes and was around 98 seconds, 4 seconds faster than when using 2,048 processes.
\textcolor{black}{Based on the presented performance characteristics and emphasizing that I/O is the dominating cost, we suggest using as many processes as required to saturate the I/O bandwidth. Increasing the dedicated resources past the I/O saturation point will significantly reduce the parallel efficiency.}

\subsection{Weak scalability} \label{sec:weak}

For the weak scalability analysis, we used the ERA5 dataset containing  the horizontal U component of wind speed on 37 pressure levels described in section~\ref{sec:geophysical}. As mentioned there, the total size of this data set is 51,745~GB, and we used between 10 and 80 years of data, corresponding to between 3.4 and 26.8 trillion data points stored using 6,273~GB to 49,863~GB of disk space. The corresponding memory required in double-precision floating-point that was adopted in this study ranged from 25,092~GB to 199,452~GB.
To maintain a constant load per process, we use 10 years of data per 2,048 processes (512 nodes) and scale up to 16,384 processes (4,096 nodes), where we use 80 years of the horizontal U component of wind speed data. Similarly to the strong scalability analysis, we split the data temporally into 1-year blocks.
\begin{figure}[H]
	\centering
    \includegraphics[width=0.72 \linewidth]{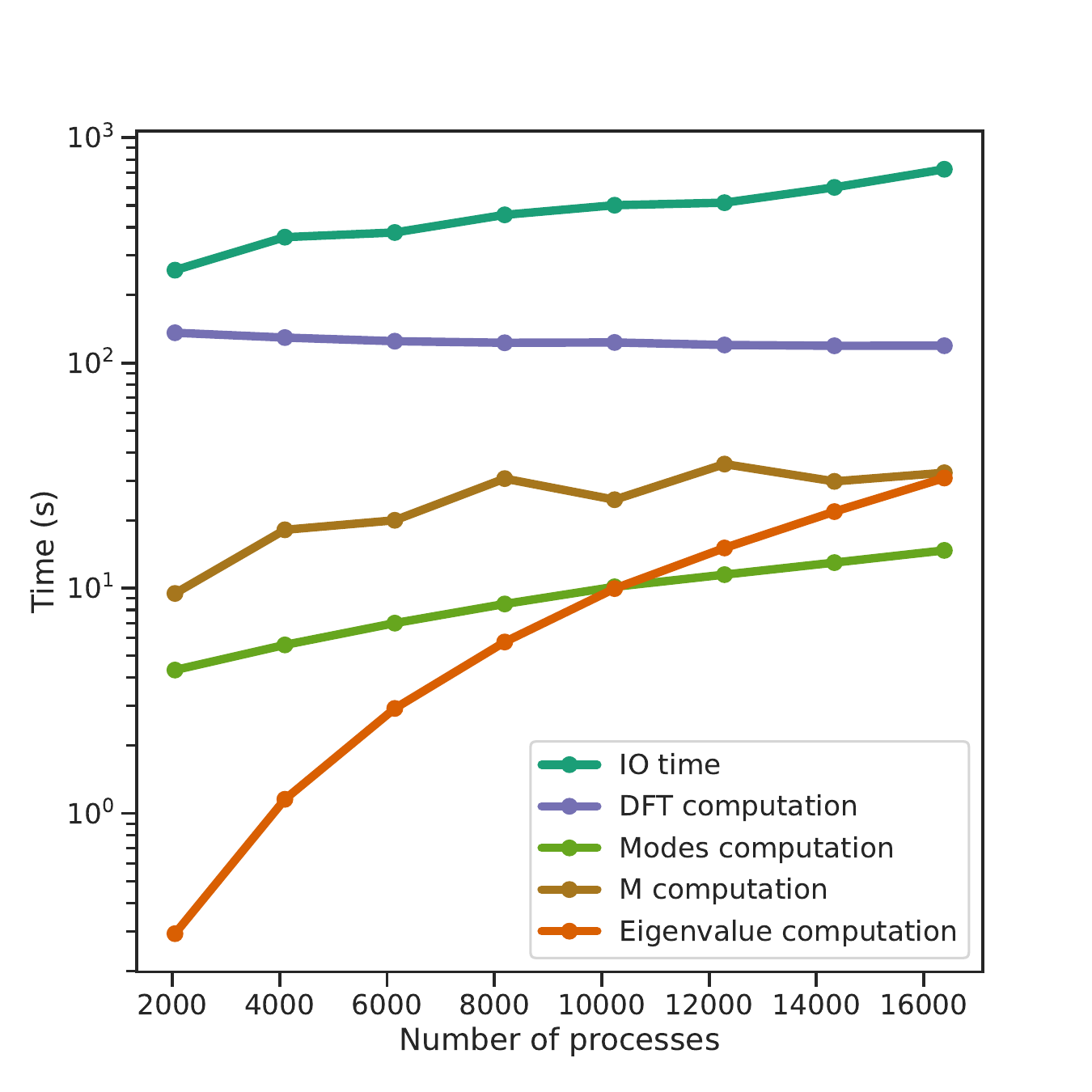}
    \caption{Weak scalability of PySPOD using the hourly horizontal speed of air data on 37 pressure levels data. 10 years of data per 2,000 processes, between January 1940 and December 2020 (when using 16,384 processes).}
    \label{fig_weak}
\end{figure}
As shown in figure~\ref{fig_weak}, we observe satisfactory I/O behavior (\textcolor{teal}{teal curve}). The bandwidth peaks at 80.41~GB/s when using 12,288 cores. DFT (\textcolor{violet}{violet curve}) and modes (\textcolor{ForestGreen}{green curve}) computation also remain efficient as more data and processes are used. However, the computation of matrix $\hat{\mathbf{M}}$ shows worse scaling and its timing displays significant variation in repeated executions (\textcolor{brown}{brown curve}). In an environment with shared interconnect, this result points to the increased collective communication cost at scale. Additionally, the eigenvalue computation (\textcolor{orange}{orange curve}), a serial component, becomes more expensive as the problem size grows.

Overall, the efficiency is 64\% when using 6,144 cores. Despite the decrease in efficiency to 38\% when using 16,384 cores, the overall runtime for this scenario is under 20 minutes, which we find acceptable given the sheer size of the dataset (49,863~GB on disk and 199,452~GB in memory).


\section{Discussion and conclusions}\label{sec:conclusions}
The new parallel SPOD algorithm allows modal decompositions that were extremely challenging if not impossible with the serial algorithms available. We were able to compute SPOD decompositions up to 199~TB using HPC platforms, and exploiting the scalability and performance of the parallel algorithm. In particular, the key novel aspect is the I/O handling dictated by the smart data layout that was devised and implemented. This allowed preserving all time operations (more specifically the DFT), and trivially distribute across MPI ranks the spatial component of the data. The results reported in section~\ref{sec:datasets} show the power of the package in providing results on big data, enabled by the scalability performance shown in section~\ref{sec:scalability}. The latter were possible thanks to an efficient implementation of I/O, that also allowed for a reduction in terms of memory consumption. The new package may allow unlocking new physics from big data that was not possible to analyze before.

\section*{Acknowledgements}
Marcin Rogowski, Lisandro Dalcin, and Matteo Parsani acknowledge support from King Abdullah University of Science and Technology (KAUST) award BAS/1/1663-01-01. Brandon C.\ Y.\ Yeung and Oliver T.\ Schmidt gratefully acknowledge support from Office of Naval Research award N00014-23-1-2457, under the supervision of Dr. Steve Martens.  Romit Maulik was supported by U.S. DOE ASCR Award Data-intensive Scientific Machine Learning: DE-FOA-2493. Gianmarco Mengaldo was supported by MOE Tier 2 grant 22-5191-A0001-0: ``Prediction-to-Mitigation with Digital Twins of the Earth’s Weather''. The authors are thankful to the KAUST Supercomputing Laboratory for their computing resources. LES calculations were carried out on the ``Onyx'' Cray XC40/50 system in ERDC DSRC, using allocations provided by DoD HPCMP.

\bibliographystyle{elsarticle-num-names}
\bibliography{references.bib}
\end{document}